\documentclass[lettersize,journal]{IEEEtran}
\usepackage{amsmath,amsfonts}
\usepackage{algorithmic}
\usepackage{algorithm}
\usepackage{array}
\usepackage{textcomp}
\usepackage{stfloats}
\usepackage{url}
\usepackage{color}
\usepackage{verbatim}
\usepackage{graphicx}
\usepackage{subcaption}
\usepackage{pifont}
\usepackage{multirow}
\usepackage{tabularx}
\usepackage{makecell}
\usepackage{colortbl}
\usepackage{xcolor}

\begin{document}

\title{Trustworthy Semantic Communication for Vehicular Networks: Challenges and Solutions}
\author{
Yanghe~Pan, Yuntao~Wang, Shaolong~Guo, Chengyu~Yin, Ruidong~Li, Zhou~Su, and Yuan~Wu
\thanks{
Y. Pan, Y. Wang, S. Guo, C. Yin, Z. Su are with the School of Cyber Science and Engineering, Xi'an Jiaotong University, Xi'an, China. (Email: zhousu@ieee.org) 

R. Li is with the College of Science and Engineering, Kanazawa University, Kanazawa, Japan. 

Y. Wu is with the Faculty of Science and Technology, University of Macau, Macau, China.
}}

\maketitle

\begin{abstract}
Semantic communication (SemCom) has the potential to significantly reduce communication delay in vehicle-to-everything (V2X) communications within vehicular networks (VNs). However, the deployment of vehicular SemCom networks (VN-SemComNets) faces critical trust challenges in information transmission, semantic encoding, and communication entity reliability. This paper proposes an innovative three-layer trustworthy VN-SemComNet architecture. Specifically, we introduce a semantic camouflage transmission mechanism leveraging defensive adversarial noise for active eavesdropping defense, a robust federated encoder-decoder training framework to mitigate encoder-decoder poisoning attacks, and an audit game-based distributed vehicle trust management mechanism to deter untrustworthy vehicles. A case study validates the effectiveness of the proposed solutions. Lastly, essential future research directions are pointed out to advance this emerging field.
\end{abstract}

\IEEEpeerreviewmaketitle

\section{Introduction}
Vehicular networks (VNs) serve as the foundation of intelligent transportation systems by enabling seamless vehicle-to-everything (V2X) connectivity. Through the interconnection of trillions of vehicles, roadside sensors, and infrastructure within smart cities, VNs support advanced applications such as intelligent traffic management, cooperative driving, and remote driving \cite{8345196}. As reported by MarketsandMarkets\footnotemark[1], the global V2X market is projected to reach USD 9.5 billion by 2030, growing at a compound annual growth rate (CAGR) of 51.9\% from 2023 to 2030.\footnotetext[1]{\url{https://www.marketsandmarkets.com/Market-Reports/automotive-vehicle-to-everything-v2x-market-90013236.html}}  
However, due to the high mobility of vehicles, direct vehicle-to-vehicle (V2V) and V2X links are often opportunistic and short-lived, making it difficult to maintain stable data exchange. 
Meanwhile, the rapid proliferation of vehicles and roadside sensors continuously generates massive multimodal perception data that needs to be timely transmitted between vehicles and roadside units (RSUs), resulting in severe bandwidth congestion and communication latency. 

Semantic communication (SemCom) has emerged as a promising solution to enhance the scalability and real-time responsiveness of V2X services by shifting the focus from transmitting raw data to conveying essential semantic information. Unlike conventional communication paradigms that prioritize bit-wise accuracy, SemCom extracts and transmits only the most relevant information, significantly reducing bandwidth usage and improving communication efficiency \cite{10798108}.  
Furthermore, the increasing onboard computational capabilities of intelligent vehicles offer opportunities to support SemCom processes such as semantic encoding and decoding. This paradigm enables low-latency, real-time V2X communication while alleviating bandwidth congestion. Additionally, SemCom facilitates context-aware information exchange, enhancing the adaptability of VNs in dynamic and complex environments. 

Recently, there are increasing works exploring the integration of SemCom and VNs. 
For instance, Xu et al. \cite{10013090} designed a multi-user collaborative SemCom architecture for VNs, optimizing semantic data selection across multiple vehicles to improve transmission efficiency. This approach outperforms conventional SemCom schemes in vehicle image retrieval tasks. 
Wan et al. \cite{10473044} proposed a multi-scene object detection method based on SemCom in VNs, where a convolutional semantic encoder transforms road image labels into semantic codes, enabling high-fidelity road image reconstruction at the receiver. 

However, the practical deployment of vehicular SemCom networks (VN-SemComNet) faces several challenges that hinder their real-world application.
1) During semantic transmission, adversaries may intercept V2V or V2X SemCom channels in VN-SemComNet, leading to eavesdropping attacks that compromise semantic confidentiality \cite{10458014}.  
2) During semantic encoding, semantic encoder-decoders in VN-SemComNet, typically trained in a federated manner across multiple vehicles, are vulnerable to poisoning and backdoor attacks \cite{10183798}. 
3) Due to the inherent heterogeneity, broad coverage, and openness of VNs, vehicles may be untrustworthy and exhibit non-cooperative behaviors, jeopardizing critical tasks such as cooperative sensing \cite{8345196}.
Hence, it is necessary to ensure the trustworthiness and reliability of VN-SemComNet in semantic transmission, semantic encoding, and communication entities. 

This paper presents an innovative three-layer trust architecture for VN-SemComNet. 
Specifically, at the semantic transmission layer, a semantic camouflage transmission mechanism is developed that incorporates defensive adversarial noise to effectively mislead potential eavesdroppers during communication. At the semantic encoding layer, a robust federated encoder-decoder training framework is devised to effectively evaluate gradient trust scores of participating vehicles and exclude poisoned gradients, ensuring robust aggregation against poisoning and backdoor attacks. Furthermore, at the communication entity layer, an audit game-based distributed vehicle trust management mechanism is proposed, enabling consensus trust construction and dynamic updates through strategic interactions between target evaluation vehicles (evaluatees) and their peer vehicles (assessors). A case study is also presented to validate the effectiveness of the proposed solutions. Finally, several future research directions in this emerging field are explored, including large model-empowered VN-SemComNet, personalized VN-SemComNet, and quantum VN-SemComNet.

\section{Overview of VN-SemComNet}

\begin{figure*}
    \centering\setlength{\abovecaptionskip}{-0.0cm}
    \includegraphics[width=1.0\textwidth]{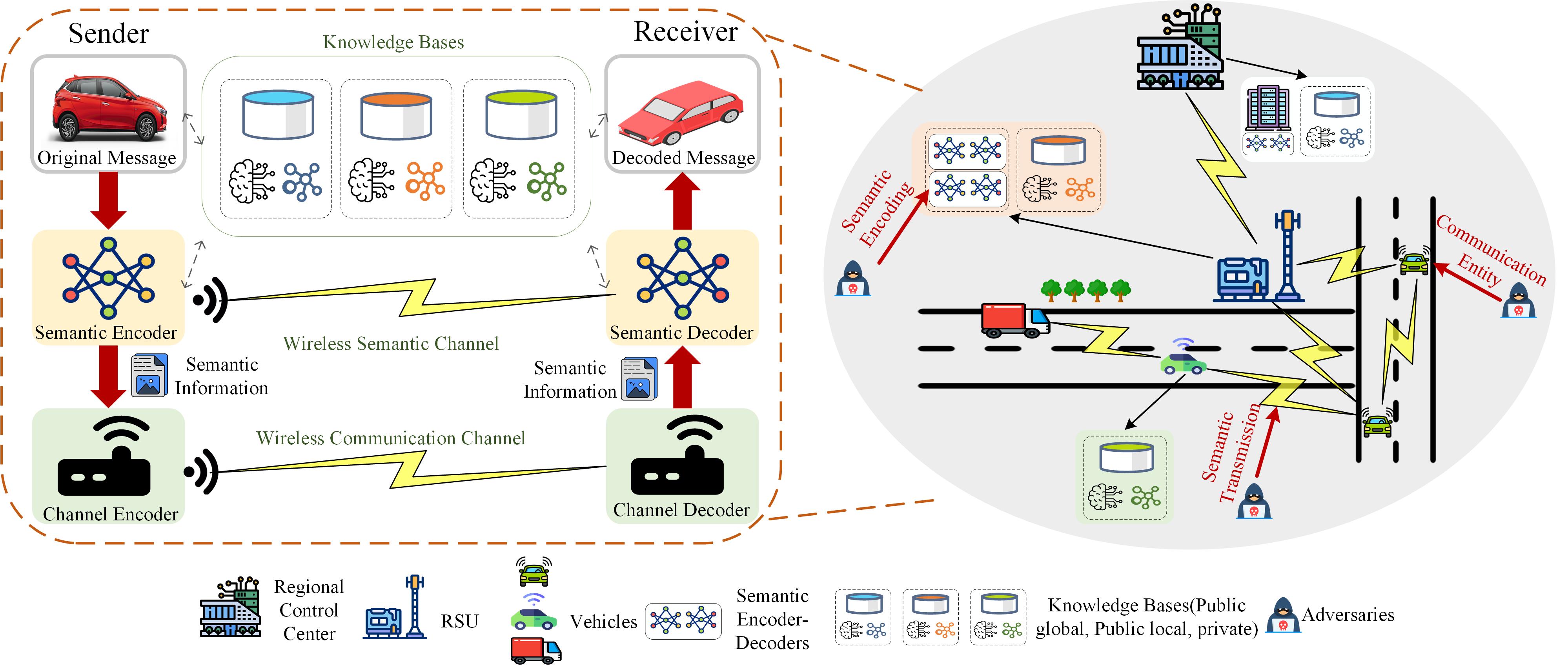}
    \caption{Overview of the general architecture of VN-SemComNet.}
    \label{fig:Overview}\vspace{-2mm}
\end{figure*}

\subsection{Architecture of VN-SemComNet}
Fig. \ref{fig:Overview} illustrates the general architecture of VN-SemComNet, comprising three primary entities: RSUs, vehicles, and regional control centers. 

\textit{Entities.} Both vehicles and RSUs are equipped with sensors, computational resources, and SemCom components, including semantic encoder-decoders, channel encoder-decoders, and knowledge bases, to support efficient SemCom. In VN-SemComNet, an RSU and vehicles within its coverage form a SemCom cluster. When a vehicle enters an RSU’s coverage, it automatically joins the cluster. Based on its requirements (e.g., target tasks and computational capabilities), the vehicle requests specific semantic encoder-decoders from the RSU, which allocates the corresponding models to enable seamless SemCom with other vehicles in the cluster.

\textit{Semantic encoder-decoders.} Semantic encoder-decoders are fundamental to VN-SemComNet. The semantic encoder extracts essential information from original data, while the decoder reconstructs the data from this information. To improve efficiency, the RSU caches frequently used encoder-decoders. If the cache cannot meet a vehicle’s requirements, the RSU obtains the required models from the regional control center and distributes them. The control center hosts a comprehensive repository of encoder-decoders based on diverse deep learning architectures, collaboratively trained via federated learning (FL) with vehicles to address varying demands across SemCom clusters.

\textit{Hierarchical knowledge bases.} VN-SemComNet adopts a three-level knowledge base architecture for efficient semantic encoding and decoding:
\begin{enumerate}
    \item \textbf{Global public knowledge base (regional control center):} Contains fundamental information, e.g., traffic regulations, maps, and driving principles, accessible to all vehicles and RSUs.
    \item \textbf{Local public knowledge base (RSUs):} Contains traffic-related information within the RSU’s coverage, including real-time conditions, incidents, and road closures, accessible to all local vehicles.
    \item \textbf{Private knolwedge base (vehicles):} Contains personal data such as perceived road conditions, driving trajectories, and vehicle status, for exclusive vehicle use.
\end{enumerate}


\textit{Channel encoder-decoders.} In VN-SemComNet, vehicles and RSUs are equipped with channel encoder-decoders. The channel encoder converts extracted semantic information into signals for physical-layer transmission, integrating redundancy mechanisms to mitigate the impact of channel noise. The channel decoder applies error correction to received signals, enabling accurate reconstruction of semantic information and ensuring reliable communication.

\textit{Advantages of VN-SemComNet.} VN-SemComNet offers advantages of low latency, efficient encoder-decoder allocation, and real-time adaptive knowledge bases. The semantic information is transmitted in VN-SemComNet directly, reducing communication bandwidth requirements. RSUs manage encoder-decoder allocation, eliminating the need for V2V encoder-decoder training between vehicles. Knowledge bases dynamically update based on real-time conditions, continually optimizing communication.

\subsection{Workflow of VN-SemComNet}
The workflow of VN-SemComNet can be divided into the following steps:
\begin{itemize}
    \item \textit{Information Sensing:} Vehicles and RSUs collect road and traffic data through various onboard and roadside sensors.
    \item \textit{Semantic Encoding:} The collected data, or the data intended for transmission, is processed by the semantic encoder at the sender to extract semantic information.
    \item \textit{Semantic Transmission:} The extracted semantic information is further encoded by the channel encoder, converting it into signals suitable for transmission over the physical communication channel.
    \item \textit{Semantic Alignment and Relaying:} When semantic alignment is required, the signal is relayed to the RSU within the SemCom cluster. The RSU decodes the signal, performs semantic correction and alignment leveraging its local public knowledge base, and forwards the refined semantic information to the intended receiver.
    \item \textit{Semantic Decoding:} The receiver applies the channel decoder to recover semantic information, then employs the semantic decoder with its private knowledge base to reconstruct the original data.
    \item \textit{Decision Making:} Based on the decoded semantic information, the receiver makes intelligent decisions such as path planning and emergency avoidance based on the decoded semantic information.
\end{itemize}

\subsection{Key Challenges of Trustworthy VN-SemComNet}
\label{subsec: challenges}
VN-SemComNet presents persistent trust challenges across multiple layers, including the information transmission layer, semantic encoding layer, and communication entity layer. 

\subsubsection{Untrustworthy Semantic Transmission}
At the information transmission layer, eavesdroppers pose a critical security threat to VN-SemComNet by joining the RSU SemCom cluster and intercepting the transmitted semantic information through V2V or vehicle-to-infrastructure (V2I) communication channels. Therefore, they can recover the original data leveraging the allocated semantic encoder-decoders from the RSU and associated knowledge bases. Conventional privacy-preserving transmission approaches such as cryptographic techniques rely on complicated key distribution, rendering them unsuitable for dynamic VN scenarios. Additionally, these approaches predominantly provide passive privacy measures and are incapable of actively misleading potential eavesdroppers.

\subsubsection{Untrustworthy Semantic Encoder-Decoders}
At the semantic encoding layer, semantic encoder-decoders are typically trained through FL between RSUs and vehicles, and are vulnerable to poisoning and backdoor attacks. Specifically, adversaries can perform model poisoning attacks \cite{fang2020local} to degrade the performance of the federated encoder-decoders by uploading noisy gradients (i.e., untargeted attacks) or manipulating the encoder-decoders to misrepresent data in specific categories (i.e., targeted attacks).

\subsubsection{Untrustworthy Vehicles}
At the communication entity layer, the inherent heterogeneity, extensive coverage, and openness of VN-SemComNet amplify the risk of untrustworthy vehicles. These vehicles may engage in non-cooperative 
or indolent activities, including providing incomplete information and delaying responses, thereby significantly undermining the reliability of VN-SemComNet and severely compromising critical tasks.
Conventional reputation evaluation schemes often struggle to accommodate the highly dynamic behavior patterns of vehicles in VN-SemComNet.

\begin{figure*}
    \centering\setlength{\abovecaptionskip}{-0.0cm}
    \includegraphics[width=1.0\textwidth]{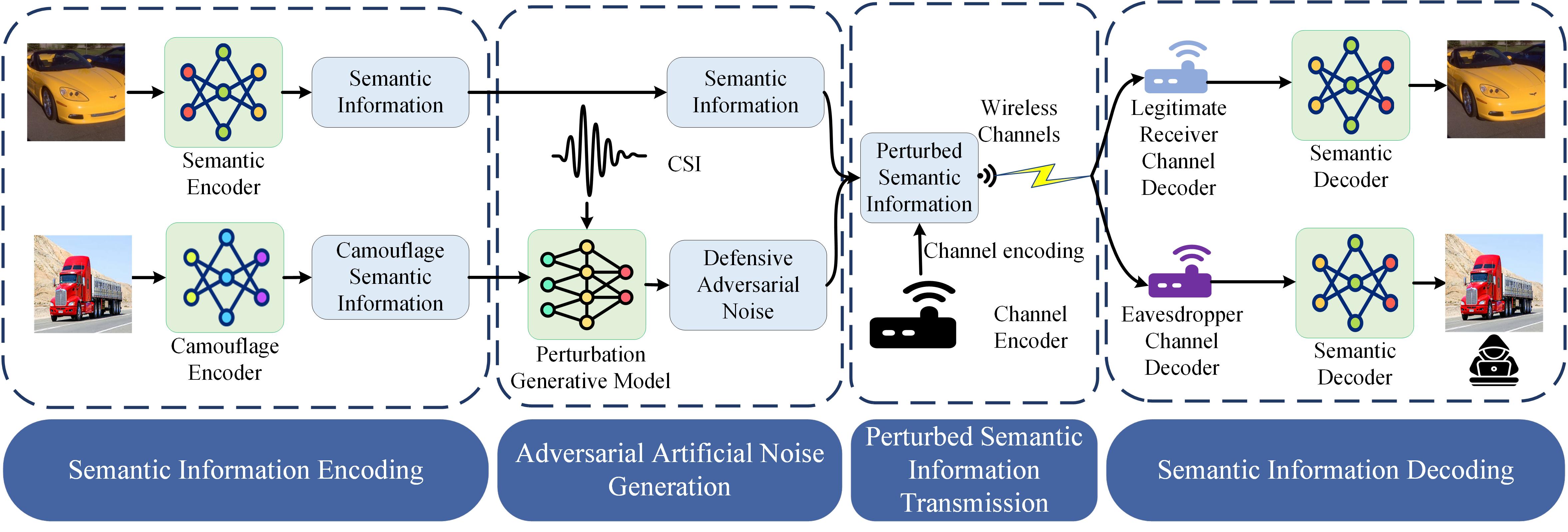}
    \caption{The workflow of semantic camouflage transmission.}
    \label{fig:semantic_camouflage_transmission}\vspace{-2.5mm}
\end{figure*}
\section{Solutions to Trustworthy VN-SemComNet}
In this section, we propose the corresponding solutions to address the key challenges of trustworthy VN-SemComNet in Sec. \ref{subsec: challenges}, including semantic camouflage transmission mechanism (Sec. \ref{subsec: adversarial noise}), reliable and trustworthy federated aggregation strategy (Sec. \ref{subsec: federated aggregation}), and distributed entity trust management mechanism (Sec. \ref{subsec: trust management}).

\subsection{Semantic Camouflage Transmission}
\label{subsec: adversarial noise}
To effectively address potential eavesdropping threats at the information transmission layer in VN-SemComNet, this subsection proposes an innovative semantic camouflage transmission approach from the perspective of physical-layer security, inspired by the concept of defensive adversarial examples \cite{frosio2023best}. {“Semantic camouflage” refers to a physical-layer security technique that perturbs encoded semantic representations in a controlled manner to mislead eavesdroppers while ensuring that legitimate users can accurately recover the intended message.} {In the proposed approach, it is assumed that the defender can obtain the channel state information (CSI) of both legitimate receivers and potential eavesdroppers through long-term observation and measurement, which is a common assumption in the research field of physical-layer security \cite{9892681}.} {The eavesdropper is assumed to have the same capability as the legitimate user, differing only in CSI. Specifically, the legitimate user’s channel includes a line-of-sight component following Rician fading, whereas the attacker’s channel lacks a line-of-sight component following Rayleigh fading.} Leveraging these CSI distinctions, the sender generates adversarial semantic-level artificial noise and superimposes it on the encoded semantic information, enabling legitimate receivers to reconstruct the original message while causing eavesdroppers to decode only the intended camouflage content.


Fig.~\ref{fig:semantic_camouflage_transmission} illustrates an example where the sender’s original message is a car image and the camouflage message is a truck image. Using the proposed defense, eavesdroppers decode only the truck image, while legitimate receivers recover the car image. 
Our approach can be divided into the following four phases:
\begin{itemize}
    \item \textit{Phase 1: Semantic information encoding.} The sender extracts the semantic information from the original message using a semantic encoder. Concurrently, an independent camouflage encoder with the same network architecture is utilized to extract the semantic information of the chosen camouflage message.
    \item \textit{Phase 2: Adversarial artificial noise generation.} The encoded camouflage semantic content is fed into an artificial noise generative model to produce adversarial semantic perturbed noise. The training data of the generative model involves the CSI of various channels, thereby enabling it to learn the distinctions among different communication channels.
    \item \textit{Phase 3: Perturbed semantic information transmission.} The sender adds the crafted adversarial noise to the encoded semantic information, producing the perturbed semantic information. After channel encoding, it is transmitted to the receiver via the VN-SemComNet wireless channel.
    \item \textit{Phase 4: Semantic information decoding.} The legitimate receiver accurately reconstructs the original message through channel and semantic decoding, while any eavesdropper can recover only the chosen camouflage message.
\end{itemize}

\subsection{Robust Federated Encoder-Decoder Training}
\label{subsec: federated aggregation}

To resist model poisoning attacks during federated encoder-decoder training at the semantic encoding layer in VN-SemComNet, this subsection proposes a robust federated training framework to ensure encoder-decoder semantic security. Initially, the aggregator (typically the regional control center) collects a standard evaluation dataset from the Internet or generates one via generative models. Subsequently, the aggregator trains the federated encoder-decoder through FL in collaboration with vehicles.


\begin{figure}
    \centering\setlength{\abovecaptionskip}{-0.0cm}
    \includegraphics[width=\linewidth]{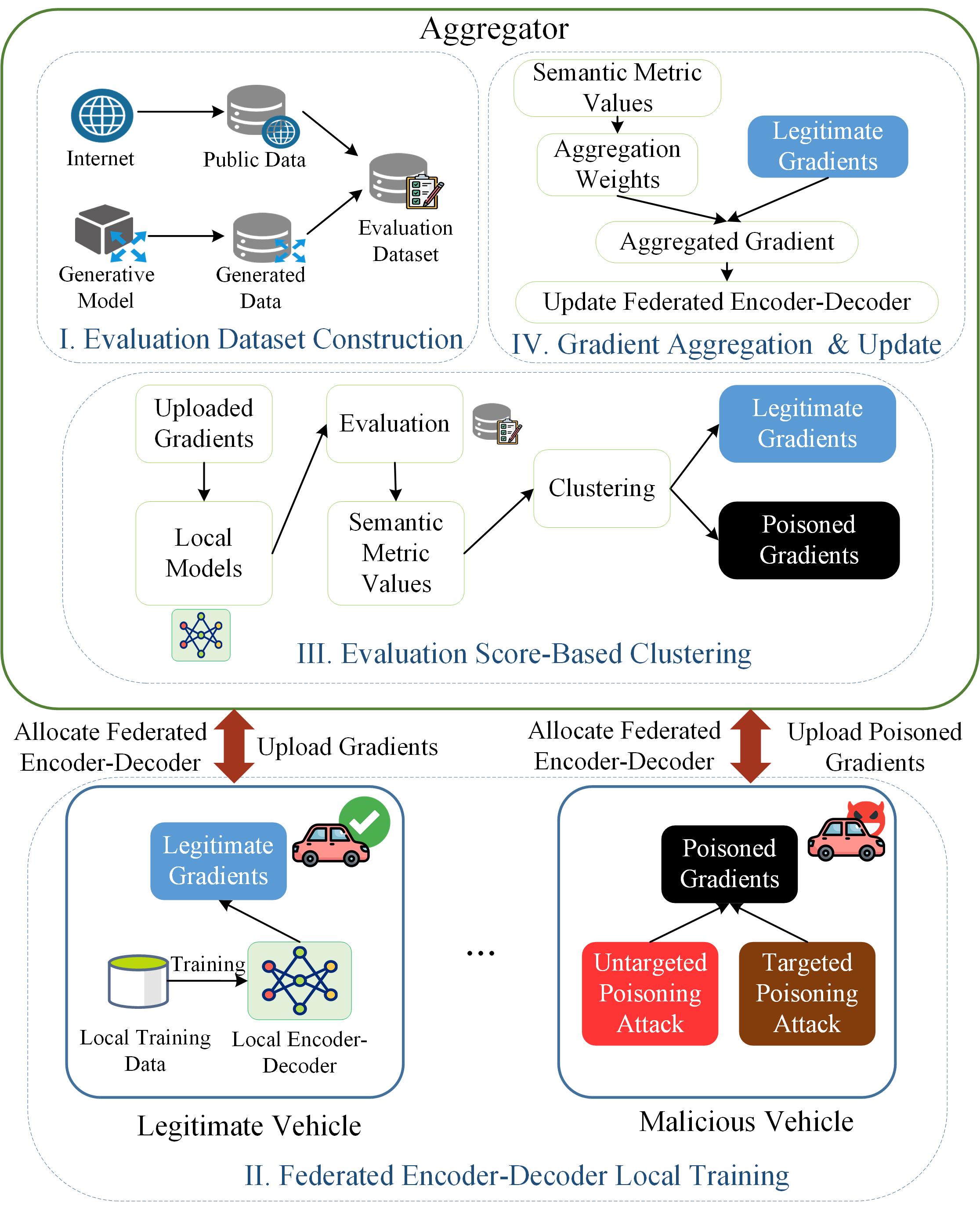}
    \caption{Illustration of robust federated encoder-decoder training framework.}
    \label{fig:aggregation}
\end{figure}

\begin{figure*}[htbp]
    \centering\setlength{\abovecaptionskip}{-0.0cm}
    \includegraphics[width=1.0\textwidth]{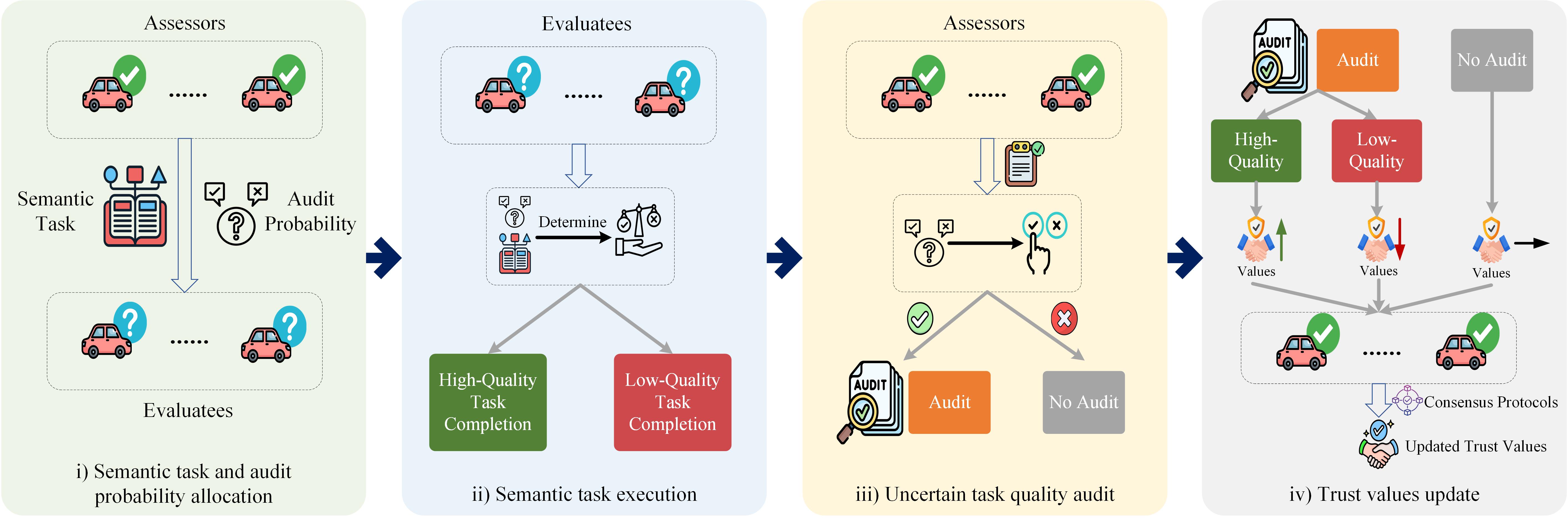}
    \caption{The workflow of audit game-based distributed vehicle trust management.}
    \label{fig: trust management}
\end{figure*}

During each FL communication round, the aggregator assesses gradients from vehicles and RSUs using the evaluation dataset, computes evaluation scores, and clusters them to exclude potential poisoning gradients. Unlike conventional similarity-based defenses \cite{DBLP:conf/ndss/CaoF0G21}, the proposed method assesses gradients based on actual model performance, better accommodating heterogeneous data across VN-SemComNet. Fig. \ref{fig:aggregation} shows the four-phase workflow of the encoder-decoder training process.

\begin{itemize}
    \item \textit{Phase 1: Evaluation dataset construction. }If sufficient public data is available, the aggregator collects data from the Internet to construct the evaluation dataset. In cases when publicly available data relevant to the SemCom within the federation is insufficient, synthetic data through generative models such as generative adversarial networks (GANs) or diffusion models can be employed to build the evaluation dataset. Both approaches can be jointly leveraged to construct the evaluation dataset. The data distribution of this dataset should closely resemble the overall distribution of local data from participating vehicles.
    \item \textit{Phase 2: Federated encoder-decoder local training.} During each communication round, the aggregator disseminates the federated encoder-decoder model to participating vehicles. Legitimate vehicles then use their local private training data to optimize the federated encoder-decoder, compute the corresponding gradients, and upload these gradients to the aggregator. Meanwhile, malicious vehicles may perform model poisoning attacks by uploading poisoned gradients to the aggregator.
    \item \textit{Phase 3: Evaluation score-based clustering. }Upon receiving the gradients uploaded by vehicles within the federation, the aggregator derives the local model for each vehicle in the current round. Subsequently, the aggregator evaluates these local models on the evaluation dataset, and calculates the semantic metric between the reconstructed information from the encoder-decoder and the original information. The aggregator then employs the clustering algorithm to evaluate score vectors, identifying and excluding potential poisoned gradients based on the clustering results.
    \item \textit{Phase 4: Trustworthy gradient aggregation and encoder-decoder update. }The aggregator allocates aggregation weights to the remaining trustworthy gradients from legitimate vehicles based on their semantic metric values. It employs a weighted aggregation algorithm to aggregate these trustworthy gradients, thereby enhancing the performance of the encoder-decoder model. The final aggregated gradient is subsequently utilized to update the federated encoder-decoder, completing the current round of federated training.
\end{itemize}

\subsection{Audit Game-Based Distributed Vehicle Trust Management}
\label{subsec: trust management}

Conventional trust management relies on centralized infrastructures and historical interaction records. Due to high vehicle mobility, it is challenging to establish trust for newly joined vehicles or quickly identify untrustworthy ones. Additionally, the reliance on centralized trust management infrastructures results in a single point of failure. To address effectively incentivize active participation of vehicles in semantic tasks and identify untrustworthy vehicles in VN-SemComNet, this subsection proposes a distributed vehicle trust management mechanism based on the audit game, deterring dishonest behavior in semantic tasks and enabling efficient trust management via uncertain audit probabilities. {Newly joined vehicles are initially treated as evaluatees and undergo trust evaluation over multiple predetermined rounds. Upon completion of these rounds, if their trust scores exceed the predefined threshold, they can become assessors.}


The workflow of the proposed mechanism can be divided into four phases, which is illustrated in Fig. \ref{fig: trust management}. \textit{i) Semantic task and audit probability allocation:} Initially, assessors (existing vehicles in VN-SemComNet) assign semantic tasks such as semantic segmentation or semantic verification tasks to evaluatees (vehicles being evaluated). Meanwhile, assessors declare the probability of auditing the quality of task completion for evaluatees. \textit{ii) Semantic task execution:} The evaluatees then determine whether to execute the assigned tasks honestly, submitting either high-quality (honest execution) or low-quality (lazy execution) results. \textit{iii) Uncertain task quality audit:} Upon receiving the results, assessors determine, based on a predefined audit probability, whether to evaluate task quality or refrain from assessment. \textit{iv) Trust values update:} Finally, assessors update the trust values of semantic evaluatees according to audit outcomes, leaving trust values unchanged if no audit occurs. These updated trust values are synchronized among assessors across VN-SemComNet using decentralized consensus protocols, ensuring global consistency and mitigating the risk of a single point of failure.

The Nash equilibrium of this audit game depends on the benefits and costs associated with task execution and quality assessment. For evaluatees, high-quality task completion incurs higher resource consumption but yields increased trust values and long-term benefits. Conversely, low-quality task completion incurs lower resource consumption but risks trust degradation if detected. For semantic assessors,  the strategic choices involve auditing or not auditing. Auditing improves trust differentiation by reinforcing high-quality evaluatees and penalizing low-quality ones, but it incurs resource costs. Conversely, the absence of such an audit conserves resources but may lead to misplaced trust, resulting in security vulnerabilities and potential losses.

\begin{figure*}[t]
\centering
\setlength{\abovecaptionskip}{0cm}

\begin{subfigure}[b]{0.48\textwidth}
    \centering
    \includegraphics[width=\linewidth]{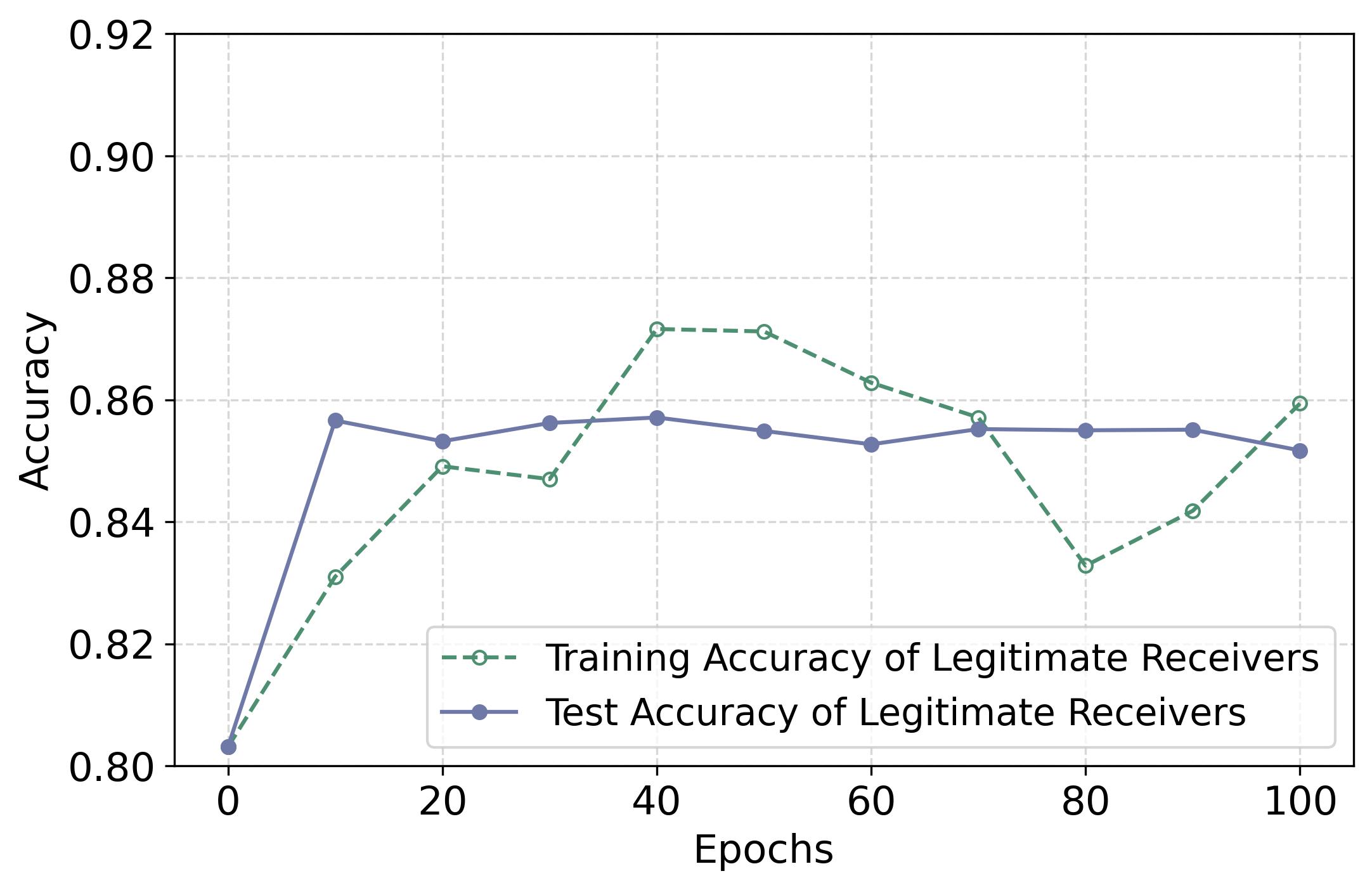}
    \caption{{Legitimate receivers' decoding performance.}}
    \label{fig:legitimate}
\end{subfigure}
\hfill
\begin{subfigure}[b]{0.48\textwidth}
    \centering
    \includegraphics[width=\linewidth]{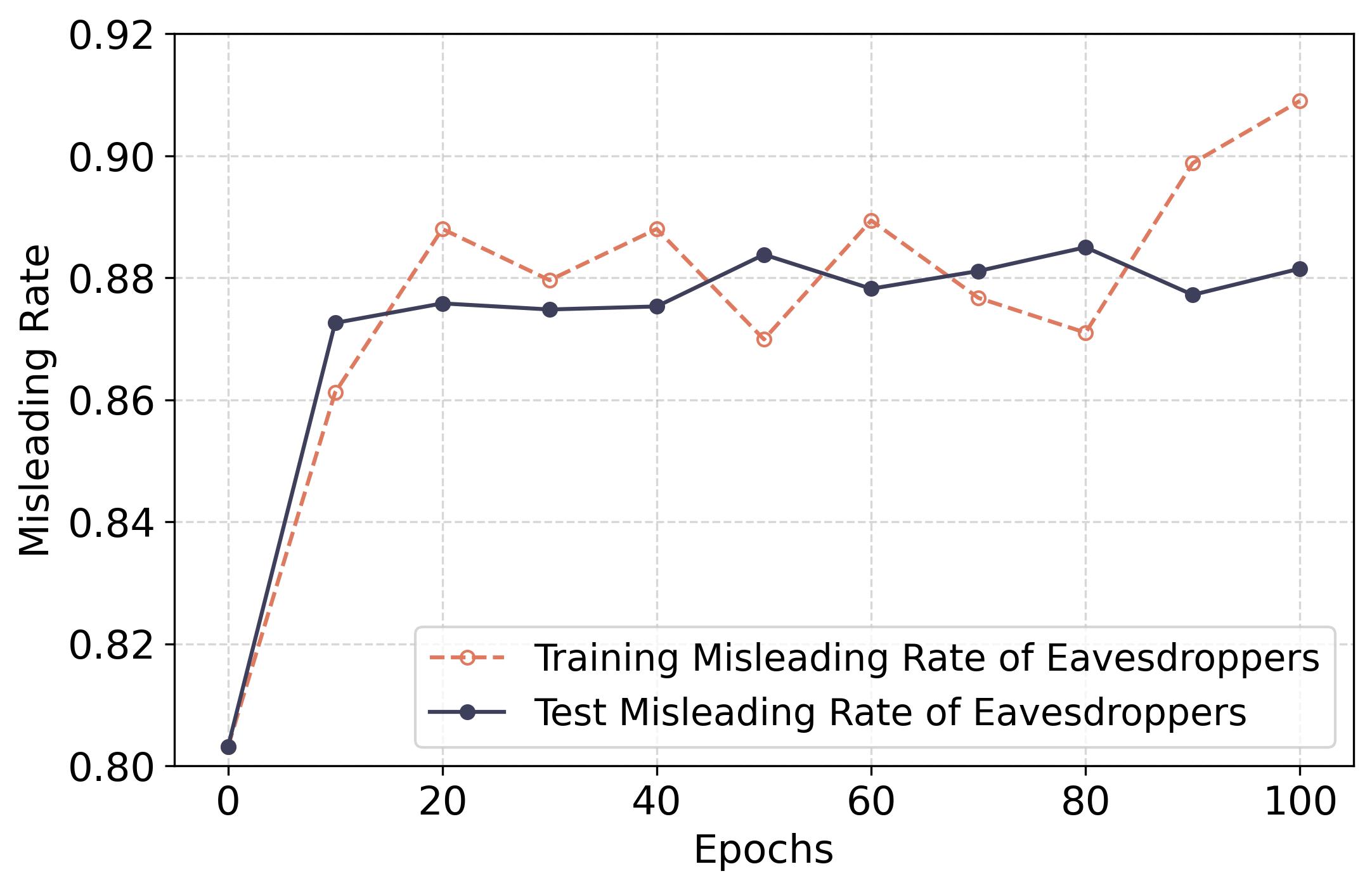}
    \caption{Semantic camouflage transmission.}
    \label{fig:misleading}
\end{subfigure}

\vspace{0.3cm} 

\begin{subfigure}[b]{0.48\textwidth}
    \centering
    \includegraphics[width=\linewidth]{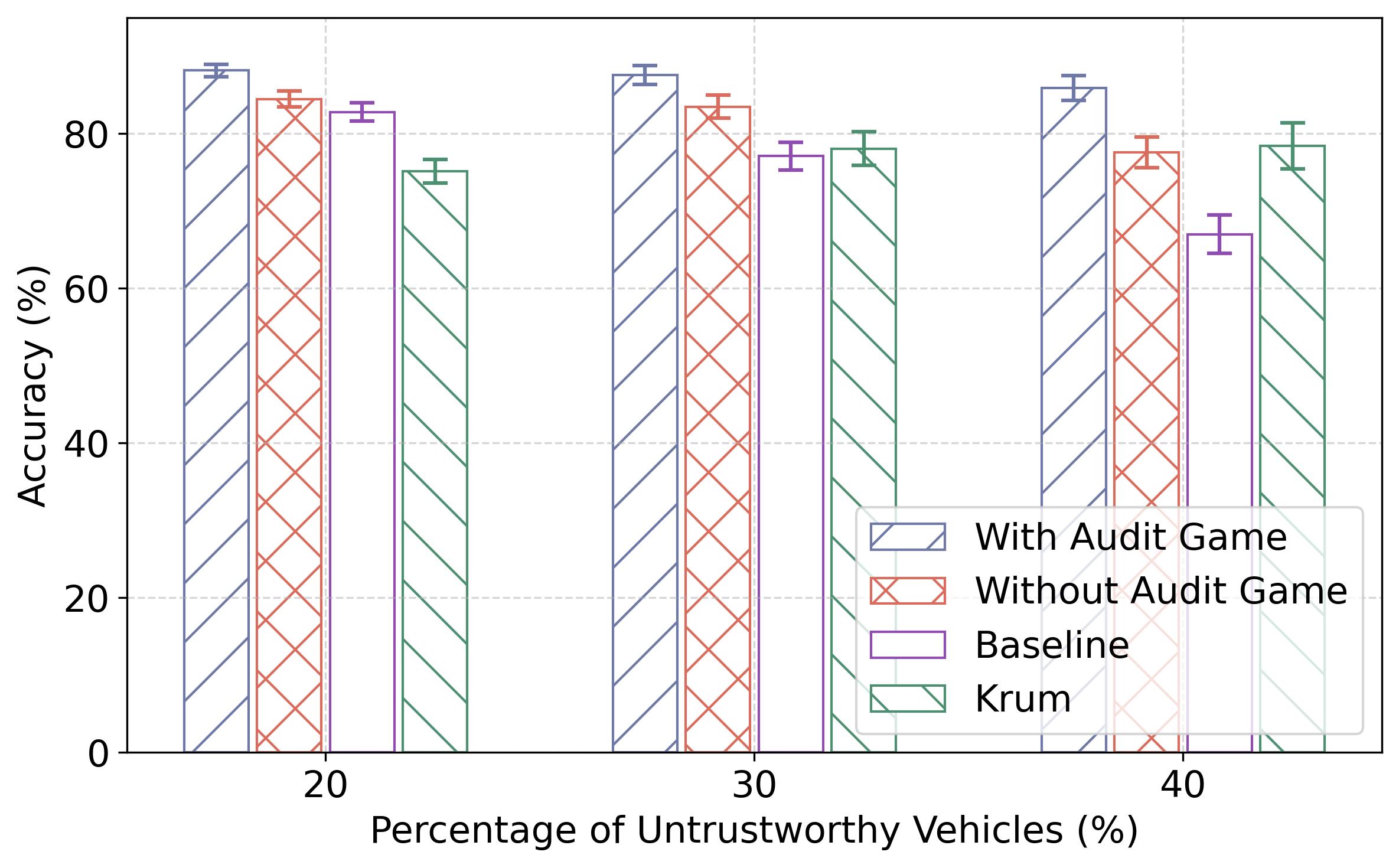}
    \caption{{Audit game-based distributed vehicle trust management.}}
    \label{fig:trust_management}
\end{subfigure}
\hfill
\begin{subfigure}[b]{0.48\textwidth}
    \centering
    \includegraphics[width=\linewidth]{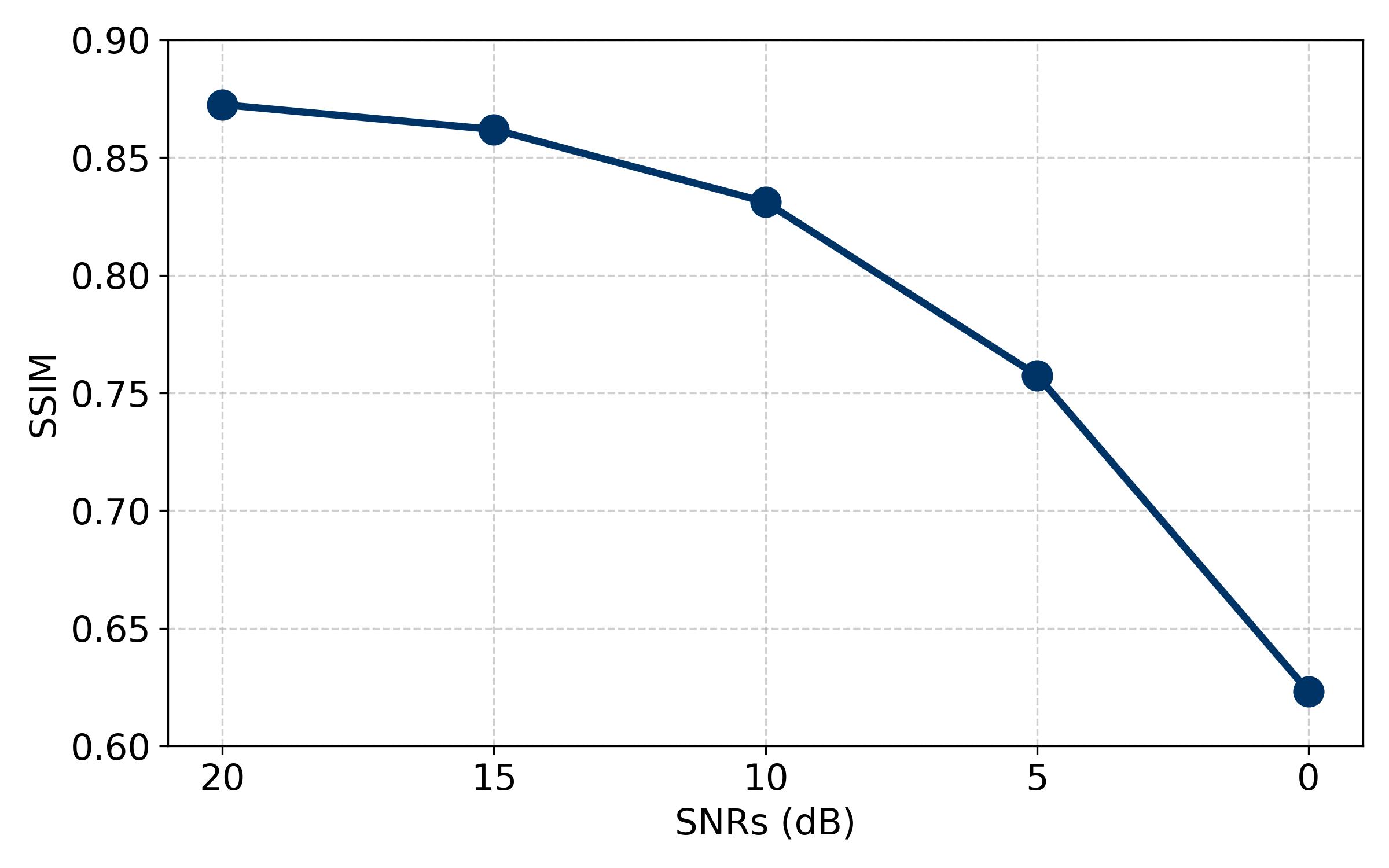}
    \caption{{Semantic communication performance under different SNRs.}}
    \label{fig:ssim_vs_snr}
\end{subfigure}

\caption{{Evaluation results of legitimate receivers' decoding performance and the misleading rate of eavesdroppers of semantic camouflage transmissions (sub-Figs (a) and (b)), the audit game-based distributed vehicle trust management (sub-Fig (c)), and the semantic communication performance under different SNRs (sub-Fig (d)).}}
\label{fig:evaluation}
\end{figure*}

\section{Case Study}
\subsection{Simulation Setup}
To evaluate semantic camouflage transmission, we utilized the semantic segmentation dataset and the encoder-decoder network architecture MFNet in \cite{8206396} to train the SemCom framework. The dataset consists of 1,569 urban scene images, segmented into nine categories. The training epochs were set to 100, and the initial learning rate was set to 0.02 with a decay rate of 0.01. The training setup of the camouflage encoder resembled that of the SemCom framework, and the noise generative model adopted the U-Net architecture \cite{falk2019u}. The misleading rate, defined as the accuracy of the result decoded by the eavesdropper concerning the camouflage image, was utilized as the metric to measure the effectiveness of semantic camouflage transmission \cite{8206396}.

To evaluate the effects of a robust federated encoder-decoder training framework, variational autoencoders (VAEs) were employed as the encoder-decoder model architecture. {The MNIST dataset and the Fashion-MNIST were utilized for performance evaluation.}

Structural similarity index measure (SSIM) was utilized as the semantic metric, effectively measuring the similarity between reconstructed and original information (e.g., images or texts) \cite{bakurov2022structural}. A higher SSIM indicates greater similarity between two messages, reflecting the quality of SemCom. {In the simulation, there were 10 vehicles for MNIST tasks and 20 vehicles for Fashion-MNIST tasks in the VN-SemComNet. The global training rounds were set to 20, with a learning rate of 1e-3.} Untrustworthy vehicles conducted model poisoning attacks (both targeted and untargeted) during the federated encoder-decoder training. {In the context of vehicular network surveillance, we evaluated communication efficiency of semantic communication in transmitting 10,000 image frames, compared with traditional byte-level transmission approaches.} 

To evaluate the impact of the audit game-based distributed vehicle trust management mechanism, we employed a federated-training semantic traffic sign recognition task utilizing GTSRB dataset. A CNN served as the model, with global accuracy on the test dataset as the evaluation metric. The network comprised 10 vehicles, with 20\%, 30\%, and 40\% designated as untrustworthy, launching adversarial attacks like label-flipping to degrade global performance. We compared the proposed mechanism with the trust management without audit game, where trust values are adjusted per communication round based on uploaded gradients and incorporated into weighted aggregation. Meanwhile, {We compare the proposed mechanism with the conventional Krum \cite{krum} mechanism}. The FL scheme without trust management serves as the baseline.

{Note that semantic communication reduces communication overhead by transmitting only task-relevant information, at the cost of additional computation for encoder-decoder training and inference. For system scalability, supporting more vehicles may require training new encoder-decoders. Moreover, vehicles with similar background knowledge can reuse the encoder-decoders to mitigate computational overhead and enahnce system scalability.}

\begin{table*}[ht]
    \centering
    \renewcommand{\arraystretch}{1.4}
    \resizebox{\textwidth}{!}{%
    \begin{tabular}{|>{\centering\arraybackslash}p{3.0cm}|*{12}{>{\centering\arraybackslash}p{1.0cm}|}}
        \hline
        \multirow{3}{*}{{\textbf{Aggregation Mechanisms}}} 
        & \multicolumn{6}{c|}{{\textbf{MNIST}}} 
        & \multicolumn{6}{c|}{{\textbf{Fashion-MNIST}}} \\
        \cline{2-13}
        & \multicolumn{3}{c|}{{\textbf{SSIM of Targeted Poisoning}}} 
        & \multicolumn{3}{c|}{{\textbf{SSIM of Untargeted Poisoning}}} 
        & \multicolumn{3}{c|}{{\textbf{SSIM of Targeted Poisoning}}} 
        & \multicolumn{3}{c|}{{\textbf{SSIM of Untargeted Poisoning}}} \\
        \cline{2-13}
        & {10\%} & {30\%} & {50\%} 
        & {10\%} & {30\%} & {50\%} 
        & {10\%} & {30\%} & {50\%} 
        & {10\%} & {30\%} & {50\%} \\
        \hline
        {FedAvg} & {0.810} & {0.783} & {0.758} & {0.564} & {0.212} & {0.117} & {0.787} & {0.770} & {0.752} & {0.630} & {0.423} & {0.289} \\
        {Proposed Architecture} & {\textbf{0.819}} & {\textbf{0.818}} & {\textbf{0.813}} & {\textbf{0.819}} & {\textbf{0.818}} & {\textbf{0.813}} & {\textbf{0.791}} & {\textbf{0.790}} & {\textbf{0.787}} & {\textbf{0.791}} & {\textbf{0.790}} & {\textbf{0.787}} \\
        \hline
        \multicolumn{13}{@{}l@{}}{\small {*Percentages indicate the proportion of adversaries.}}
    \end{tabular}%
    }
    \caption{{Average SSIM of the proposed architecture and conventional FedAvg mechanism under model poisoning attacks on MNIST and Fashion-MNIST, given different proportions of untrustworthy vehicles. Higher SSIM values indicate stronger performance of the semantic encoder-decoder.}}
    \label{tab:ssim_comparison}
\end{table*}

\subsection{Evaluation Results}

{Fig.~\ref{fig:evaluation}(a) and (b) show the decoding performance of legitimate receivers and misleading rates of eavesdroppers during training, evaluated on both training and test datasets. In Fig.~\ref{fig:evaluation}(a), both the classification accuracy on training and test sets converge quickly and remain stable.} Fig.~\ref{fig:evaluation}(b) shows misleading rates with red dotted (training) and black solid (test) lines, reaching 88\% and 91\% on the test set. The results demonstrate that the semantic camouflage mechanism can effectively preserve user privacy in VN-SemComNet. {Fig.~\ref{fig:evaluation}(c) compares global model accuracy among trust mechanisms with audit game, trust mechanism without audit game, Krum mechanism, and the baseline without trust management.} The results show that the proposed mechanism consistently outperforms the other approaches across all percentages of untrustworthy vehicles. {Fig.~\ref{fig:evaluation}(d) illustrates semantic communication performance under varying SNR levels. SSIM remains high with only a gradual decline from 20 dB to 5 dB and a moderate drop at 0 dB, demonstrating the robustness of our approach against severe channel degradation and transmission reliability under poor vehicular communication conditions.}

{Table~\ref{tab:ssim_comparison} compares SSIM between the proposed approach and FedAvg \cite{mcmahan2017communication} under targeted and untargeted poisoning attacks on MNIST and Fashion-MNIST. For MNIST targeted attacks, FedAvg’s SSIM drops from 0.810 to 0.758 as adversaries increase from 10\% to 50\%, while our approach maintains between 0.819 and 0.813. Under untargeted attacks, FedAvg's SSIM falls sharply from 0.564 to 0.117, but the proposed approach remains stable.} {Similar trends appear on Fashion-MNIST, underscoring the robustness of the proposed VN-SemComNet in preserving semantic encoding integrity.}

{In the vehicular network surveillance scenario, compared to the traditional raw data transmission approach, semantic communication reduces latency and overhead by approximately 19.6× and 19.55×, respectively. It demonstrates a substantial improvement in transmission efficiency while preserving the integrity of conveyed semantic information.}

\section{Future Research Directions}

\subsection{Large Model Empowered VN-SemComNet}
The effectiveness of VN-SemComNet depends on comprehensive, diverse, and high-quality knowledge bases. Large models, with vast parameters and extensive background knowledge, provide a universal approach to constructing and enriching these bases. They generate precise semantic representations across scenarios and can be fine-tuned efficiently. However, large models remain vulnerable to security and privacy threats, especially \textit{hallucinations}, which may generate incorrect or irrelevant content, undermining VN-SemComNet’s trustworthiness. 

\subsection{Personalized VN-SemComNet}
A personalized VN-SemComNet tailors the SemCom process based on a vehicle’s profile, preferences, and context, significantly enhancing the quality of experience by ensuring transmitted information is relevant and comprehensible. Machine learning–powered dynamic profiling enables real-time adaptation of communication strategies. However, privacy concerns arise from analyzing private behaviors and preferences. Integrating privacy-preserving technologies like federated learning with dynamic entity analysis to develop personalized VN-SemComNet remains a promising direction.

\subsection{Quantum VN-SemComNet}
Quantum SemCom leverages quantum principles to enhance the efficiency and security of semantic transmission. Quantum computing enables high-speed processing, while quantum communication techniques such as quantum entanglement and superposition can transmit vast semantic information simultaneously. Moreover, quantum cryptography addresses privacy and security threats like eavesdropping and tampering. However, practical quantum VN-SemComNet faces challenges due to a lack of quantum-compatible semantic models, efficient processing algorithms, and communication infrastructure.

\section{Conclusion}
In this paper, we introduce an innovative VN-SemComNet architecture, as well as key challenges in designing trustworthy VN-SemComNet. To address these challenges, we propose three solutions spanning the information transmission, semantic encoding, and communication entity layers. The solutions include a semantic camouflage transmission approach using defensive adversarial noise for active eavesdropping defense, a robust federated encoder-decoder training framework to resist encoder-decoder poisoning attacks, and a signal game-based distributed vehicle trust management mechanism to filter out untrustworthy vehicles. A case study is conducted to validate the effectiveness of the proposed solutions, followed by a discussion on key future research directions for future VN-SemComNet design.

\bibliographystyle{IEEEtran}  
\bibliography{bib_reference}

\end{document}